\documentclass[floats,floatfix,showpacs,amssymb,prd,superscriptaddress,nofootinbib,twocolumn,aps,eqsecnum]{revtex4-1}

\usepackage[utf8]{inputenc}

\usepackage{amsmath, amsthm, amsfonts, amssymb, latexsym}
\usepackage{graphicx}
\usepackage{url}

\def\rbl{r_\mathrm{BL}}
\def\rar{r_\mathrm{areal}}

\usepackage[unicode]{hyperref}
\hypersetup{
    unicode=true,          
    pdftitle={Testing the nonlinear stability of Kerr-Newman black holes},
    pdfsubject={Numerical Relativity},
    pdfcreator={LaTeX},          
    pdfproducer={LaTeX},         
    pdfkeywords={General Relativity, Numerical Relativity, Black Holes},
    colorlinks=false,            
  }

\begin{document}

\title{Testing the nonlinear stability of Kerr-Newman black holes}

  \author{Miguel~Zilh\~ao}
  \affiliation{
    Center for Computational Relativity and Gravitation and School of Mathematical Sciences, 
    Rochester Institute of Technology, 
    Rochester, NY 14623, USA
  }
  \affiliation{
    Departament de F\'{\i}sica Fonamental \& Institut de Ci\`{e}ncies del Cosmos, 
    Universitat de Barcelona, 
    Mart\'{\i} i Franqu\`{e}s 1, E-08028 Barcelona, Spain
  }

  \author{Vitor~Cardoso}
  \affiliation{
    CENTRA, Departamento de F\'{\i}sica, Instituto Superior T\'ecnico, Universidade de Lisboa,
    Avenida Rovisco Pais 1, 1049 Lisboa, Portugal.
  }
  \affiliation{
    Perimeter Institute for Theoretical Physics, Waterloo, Ontario N2L 2Y5, Canada
  }

  \author{Carlos~Herdeiro}
  \affiliation{
    Departamento de F\'\i sica da Universidade de Aveiro and I3N, 
    Campus de Santiago, 3810-183 Aveiro, Portugal
  }
  
  \author{Luis~Lehner}
  \affiliation{
    Perimeter Institute for Theoretical Physics, Waterloo, Ontario N2L 2Y5, Canada
  }

  \author{Ulrich~Sperhake}
  \affiliation{
    Department of Applied Mathematics and Theoretical Physics,
    Centre for Mathematical Sciences, University of Cambridge,
    Cambridge CB3 0WA, UK
  }
  \affiliation{
    Theoretical Astrophysics 350-17,
    California Institute of Technology,
    Pasadena, CA 91125, USA
  }
  \affiliation{
    Department of Physics and Astronomy,
    The University of Mississippi,
    University, MS 38677, USA
  }


\begin{abstract}
The nonlinear stability of Kerr-Newman black holes (KNBHs) is investigated by performing numerical simulations within the full Einstein-Maxwell theory. We take as initial data a KNBH with mass $M$, angular momentum to mass ratio $a$ and charge $Q$. Evolutions are performed to scan this parameter space within the intervals  $0\le a/M\le 0.994$ and $0\le Q/M\le 0.996$, corresponding to an extremality parameter $a/a_{\rm max}$ ($a_{\rm max} \equiv \sqrt{M^2-Q^2}$) ranging from $0$ to $0.995$. These KNBHs are evolved, together with a small bar-mode perturbation, up to a time of order $120M$. Our results suggest that for small $Q/a$, the quadrupolar oscillation modes depend solely on $a/a_{\rm max}$, a universality also apparent in previous perturbative studies in the regime of small rotation. Using as a stability criterion the absence of significant relative variations in the horizon areal radius and BH spin, we find no evidence for any developing instability. 
\end{abstract}


\maketitle


\section{Introduction}

According to the celebrated uniqueness theorems (see~\cite{Robinsoon:2004zz,Chrusciel:2012jk} for reviews), the Kerr-Newman (KN) solution~\cite{Newman:1965my} describes the most general  stationary, regular (on and outside a horizon) single black hole (BH) configuration of Einstein-Maxwell theory. The solution is a 4-parameter family, described by mass $M$, angular momentum $J$, electric charge $Q$ and magnetic charge $P$. The magnetic charge, however, besides being absent in standard electrodynamics, can be removed by using the electromagnetic duality of the electrovacuum Einstein-Maxwell theory~\cite{Deser:1976iy}. As such, it is often neglected---as will be the case here.

Even though it is unlikely that the KNBH plays a relevant role in astrophysics~\cite{Gibbons:1975kk,Blandford:1977ds,Barausse:2014tra}, this solution has raised considerable interest since its discovery, as an arena for theoretical investigations. In particular, it provides an ideal testing ground for studying the interplay between gravity and electrodynamics at a nonlinear level and the extent to which fundamental properties of the Kerr space-time are modified by the electromagnetic field.

As for similarities, the KN line element is of course remarkably similar to that of the Kerr solution. 
In particular, special properties of Kerr also apply to the
more general KN spacetime. For instance, the Liouville integrability of the geodesic equations observed in Kerr, is still present in KN~\cite{Carter:1968rr}. Indeed, KN possesses a hidden constant of motion, which permits the separability of test particle equations. Geometrically, this conserved quantity can be understood from the existence of an irreducible Killing tensor~\cite{Carter:1977pq}. Yet another consequence of this hidden symmetry is that scalar perturbations, obtained by solving the scalar wave equation in the KN background, are separable~\cite{Carter:1968ks}.

A different behaviour, on the other hand, is found when considering electromagnetic and gravitational perturbations. Electromagnetic fluctuations are decoupled from gravitational fluctuations in the Kerr geometry, and both separate in an elegant way when using the Newman Penrose formalism~\cite{Teukolsky:1972my}. These properties allowed for a number of significant results to be achieved for the Kerr geometry, most notably its mode stability~\cite{Whiting:1988vc}. In contrast, electromagnetic and gravitational perturbations do not decouple in the KN background
and need to be studied jointly: a small gravitational fluctuation in such background induces a perturbation in the electromagnetic field which is of the same order of magnitude. The separability of the relevant equations in the KNBH background is a formidable open problem~\cite{MTB}. This difficulty has prevented the analysis of various physical properties of the KNBH, most notably its mode stability and oscillation properties.

Understanding the stability of a solution to Einstein's equations plays a central role in assessing the solution's physical relevance. As such, considerable effort has been devoted towards establishing a proof of the stability of the Kerr solution beyond mode  analysis~\cite{Dafermos:2010hd}. Also, mounting (but certainly partial) evidence for stability has been
furnished by a large body of numerical simulations performed over the last decades. These include
binary mergers of BHs and/or neutron stars as well as rotating stars undergoing collapse. These efforts have accumulated considerable support for this solution being stable at the nonlinear level as well, at least within the 
time-scales and regimes probed by these simulations (we refer the reader to~\cite{Lehner:2014asa} for a further
discussion of this point and numerous representative references of relevant examples, and to~\cite{Cardoso:2014uka} for similar efforts in a broader context).

Much less is known about the stability of the KN solution. Indeed, due to the difficulties mentioned above, progress has only been made recently in studying electromagnetic and gravitational perturbations in either the slow or extreme rotation limits. For instance, thorough analysis of the behavior of perturbations in the slow rotation regime
indicate that KN is linearly stable for all values of the charge~\cite{Pani:2013ija,Pani:2013wsa}. The methods used for this purpose, however, are not able to probe fast rotating KN solutions and thus require a different strategy. In recent years, interesting perturbative approaches have
been developed that exploit the particular structure arising in the near-extreme limit either directly~\cite{Yang:2013uba,Yang:2014tla} 
or through the Kerr/CFT correspondence and the expanded set of isometries arising in such scenarios~\cite{Porfyriadis:2014fja,Hadar:2014dpa}. These works
are providing incipient evidence for linear stability in near-extremal BHs.
In this work we shall explore the nonlinear stability of the KN solution using tools from numerical relativity, which allow us to probe the fast rotating limit (see also~\cite{East:2013mfa} for studies in the non-charged case). 

The formalism we employ here has been described in Refs.~\cite{Zilhao:2012gp,Zilhao:2013nda}, which was previously
employed to study collisions of charged BHs with equal and with opposite charges in Einstein-Maxwell theory. With this formalism we are able, after choosing appropriate initial data, to analyse the behavior of perturbed KNBHs. 
Anticipating some of the discussions, we will show that our evolutions reveal no evidence for instabilities. 

This paper is organized as follows. In Sec.~\ref{sec:evol-eq} we briefly review the formalism used in~\cite{Zilhao:2012gp,Zilhao:2013nda} for evolving the Einstein-Maxwell system. Section~\ref{sec:init-data} addresses the construction of appropriate initial data to describe a (perturbed) KNBH. Section~\ref{sec_diagnosis} describes the diagnostic tools used to monitor the evolution and decide on whether instabilities are present. The numerical results are reported in Sec.~\ref{sec:numerical_results} and our conclusions and final remarks are made in Sec.~\ref{sec:final}.

\section{Formalism}
\label{sec:evol-eq}

Following our previous work on collisions of charged BHs in Refs.~\cite{Zilhao:2012gp,Zilhao:2013nda}, we consider the enlarged electrovacuum Einstein-Maxwell equations
\begin{equation}
  \label{eq:EFE}
  \begin{aligned}
    R_{\mu \nu} - \frac{R}{2} g_{\mu \nu} & = 8\pi T_{\mu \nu} \  ,\\
    \nabla_{\mu}\left( F^{\mu \nu} + g^{\mu\nu} \Psi
    \right) & = -\kappa n^{\nu} \Psi \ , \\
    \nabla_{\mu}\left(
    \star \!{}F^{\mu \nu} + g^{\mu\nu} \Phi
    \right) & = -\kappa n^{\nu} \Phi \ ,
  \end{aligned}
\end{equation}
where $F_{\mu \nu}=\partial_{\mu}A_{\nu} - \partial_{\nu}A_{\mu}$ is the Maxwell tensor and $\star \!{}F^{\mu \nu}$ its Hodge dual, $\kappa$ is a constant and
$n^\mu$ is the 4-velocity of Eulerian observers. We recover the
standard Einstein-Maxwell system when $\Psi = 0 = \Phi$ and merely introduce these fields as a means to damp and control violations of the
magnetic and electric constraints during the numerical evolution~\cite{Komissarov:2007wk,Palenzuela:2008sf}.
The electromagnetic stress-energy tensor takes the usual form
\begin{equation}
  \label{eq:Tmunu}
  T_{\mu \nu} = \frac{1}{4\pi} \left[ F_{\mu}{}^{\lambda} F_{\nu \lambda}
    - \frac{1}{4} g_{\mu \nu} F^{\lambda \sigma} F_{\lambda \sigma}
    \right] \ .
\end{equation}

We perform a Cauchy (3+1) decomposition by introducing a 3-metric
  $\gamma_{\mu\nu} = g_{\mu \nu} + n_{\mu} n_{\nu}$,
and decompose the Maxwell tensor and its dual into the electric and magnetic 4-vectors as
\begin{equation}
  \begin{aligned}
    F_{\mu \nu} & = n_{\mu} E_{\nu} - n_{\nu} E_{\mu}
      + \epsilon_{\mu\nu\alpha\beta} B^{\alpha} n^{\beta}  \ ,\\
      \star \! F_{\mu \nu} & = n_{\mu} B_{\nu} - n_{\nu} B_{\mu}
      - \epsilon_{\mu\nu\alpha\beta} E^{\alpha} n^{\beta}  \ ,
  \end{aligned}
  \label{eq:faraday}
\end{equation}
where we use the convention $\epsilon_{1230} = \sqrt{-g}$,
$\epsilon_{\alpha \beta \gamma}
= \epsilon_{\alpha \beta \gamma \delta} n^{\delta}$,
$\epsilon_{123} = \sqrt{\gamma}$.

\section{Initial data}
\label{sec:init-data}

As already mentioned in the Introduction, the KN solution is defined by three (physical) parameters: mass $M$, spin $a M$ and electric charge $Q$.
In Boyer-Lindquist coordinates $(t,\rbl,\theta,\phi)$, the metric and vector potential take the form (see, e.g., Ref.~\cite{wald1984general})
\begin{equation}
\begin{aligned}
  ds^2 & = - \left( \frac{\Delta - a^2 \sin^2 \theta}{\rho^2} \right) dt^2
             + \frac{\rho^2}{\Delta} d\rbl^2 
             +  \rho^2 d\theta^2  \\
           & \quad + \frac{ (\rbl^2 + a^2)^2 
             - \Delta a^2 \sin^2 \theta}{\rho^2} \sin^2 \theta d\phi^2 \\
           & \quad - \frac{2a \sin^2 \theta(\rbl^2 + a^2 - \Delta)}{\rho^2} dt d\phi \,, \\
\mathcal{A} & = -\frac{Q \rbl}{\rho^2}\left(dt-a \sin^2 \theta d\phi \right) \,,
\end{aligned}
  \label{eq:KN-BL-ds}
\end{equation}
where
\begin{align*}
\rho^2 & \equiv \rbl^2 + a^2 \cos^2 \theta \,, \\
\Delta & \equiv \rbl^2 - 2 M \rbl + a^2 + Q^2 \,.
\end{align*}

In order to obtain initial data suitable for numerical evolutions using the
``moving punctures'' technique \cite{Campanelli:2005dd,Baker:2005vv},
we express the solution in terms of a quasi-isotropic radial coordinate $R$. Following Refs.~\cite{Brandt:1996si,Krivan:1998td,Cook:2000vr} we perform the coordinate transformation
\begin{equation}
\rbl = R \left(1+\frac{M + \sqrt{a^2 + Q^2}}{2R}\right)
\left(1 + \frac{M - \sqrt{a^2 + Q^2}}{2R}\right)\,,\nonumber
\end{equation}
and the metric then takes the form [$\bar i, \bar j = (R,\theta,\phi)$ are spatial indices]
\begin{equation}
\label{eq:KN-3+1}
ds^2 = \left(-\alpha^2 + \beta_{\phi}\beta^{\phi}
\right) dt^2 + 2 \beta_\phi d\phi dt+ \gamma_{\bar i \bar j} dx^{\bar i} dx^{\bar j} \,,
\end{equation}
where
%
\begin{align*}
\gamma_{\bar i \bar j} dx^{\bar i} dx^{\bar j} &= 
\psi^4 \Big[dR^2 + R^2 \left( d\theta^2 + \sin^2 \theta d\phi^2 \right) \\
& \qquad + a^2 h R^4 \sin^4\theta d\phi^2\Big] \,, \\
\alpha & = \frac{\left(R+R_H \right) \left(R-R_H \right)}
{R \sqrt{\rbl^2 + a^2(1+\sigma \sin^2 \theta)} } \,, \\
\beta_{\phi}  &= - a \sigma \sin^2 \theta \,, \qquad \beta^{\phi}=\beta_{\phi}/\gamma_{\phi \phi} \,, \\
\psi^{4} & \equiv \rho^2/R^2, \qquad
h \equiv (1+\sigma)/(\rho^2 R^2), \\
\sigma & \equiv (2M \rbl - Q^2)/\rho^2.
\end{align*}
%
Here, $R_H \equiv \frac{1}{2} \sqrt{M^2-a^2-Q^2}$ is the location of the event horizon in the quasi-isotropic coordinate $R$. 
The nonzero components of the extrinsic curvature $K_{\bar i \bar j}$ take the form
\begin{equation}
K_{R \phi}  = \psi^{-2} \frac{H_E \sin^2 \theta}{R^2} \,,\quad K_{\theta \phi} = \psi^{-2} \frac{H_F \sin \theta}{R} \,, 
\end{equation}
where
\begin{equation}
\label{eq:aux}
\begin{aligned}
H_E & \equiv \frac{a M G}{\rho^3 \sqrt{\rbl^2 + a^2 (1 + \sigma \sin^2 \theta)} } \,, \\
H_F & \equiv -\alpha \sigma a^3 \frac{\cos\theta \sin^2 \theta}{\rho} \,, \\
G   & \equiv (\rbl^2 - a^2)\rho^2 + 2\rbl^2 (\rbl^2 + a^2) \\
    & \quad - \frac{Q^2}{M} \rbl \left( 2 \rho^2 + a^2 \sin^2 \theta 
              \right) \, .
\end{aligned}
\end{equation}

The electric and magnetic field can be computed from~(\ref{eq:faraday}). Its nonzero components are
\begin{equation}
  \begin{aligned}
    E^R & = \frac{Q R (2\rbl^2 - \rho^2) (\rbl^2 + a^2)}
    {\rho^6 \sqrt{\rbl^2 + a^2 (1 + \sigma \sin^2 \theta ) } } \,, \\
    E^{\theta} & = - \frac{2 a^2 Q \alpha R \cos\theta \sin\theta}{\rho^6}  \,, \\
    B^R & = \frac{2 a Q R \rbl (\rbl^2 + a^2) \cos \theta }
             {\rho^6 \sqrt{\rbl^2 + a^2 (1 + \sigma \sin^2 \theta ) } } \,, \\
    B^{\theta} & = \frac{a Q \alpha (2 \rbl^2 - \rho^2) \sin\theta}{\rho^6}  \,.
\end{aligned}\label{eq:EB}
\end{equation}

We finally transform to Cartesian coordinates $ x^{\bar i} = (t,R,\theta,\phi) \to x^{i} = (t,x,y,z)$. Our initial data then reads
\begin{align}
& \gamma_{ij} dx^i dx^j = \psi^4 \big[ 
     dx^2 + dy^2 + dz^2 \notag \\
 & \qquad \qquad \quad  + a^2 h \left( y^2 dx^2 - 2xy dx dy + x^2 dy^2 \right)
                            \big] \,,\label{eq:gamma0ij} \\
& K_{ij} = \Lambda^{\bar i}{}_{i} \Lambda^{\bar j}{}_{j} K_{\bar i \bar j} \,, \quad
E^{i} = \Lambda^{i}{}_{\bar i} E^{\bar i} \,, \quad
B^{i} = \Lambda^{i}{}_{\bar i} B^{\bar i} \,,
\end{align}
%
where $\Lambda^{\bar i}{}_{i} = \frac{\partial x^{\bar i}}{\partial x^{i}} $, in a form analogous to that of Refs.~\cite{Hannam:2006zt,Shibata:2009ad,Shibata:2010wz}.

In order to study the stability of this solution, we follow~\cite{Shibata:2009ad,Shibata:2010wz} and introduce a small bar-mode perturbation to the 3-metric $\gamma_{ij}$ and specify the initial conditions for the 3-metric elements as
\begin{equation}
\hat \gamma_{ij} = \gamma_{ij} \left[
 1 + A \frac{x^2 - y^2}{M^2} e^{-\frac{(R-R_0)^2}{2R_H^2}}
\right]^{-1} \,, \label{eq:pert}
\end{equation}
where $A \ll 1$, $\gamma_{ij}$ is the unperturbed solution given by~\eqref{eq:gamma0ij}, and $R_0$ is a tunable parameter that localizes the perturbation.


This perturbation is constraint violating\footnote{Constraint violations are an
  inherent consequence of the numerical modelling of spacetimes in general
  relativity at the level of the numerical discretization error. Following a
  common approach (see e.g.~\cite{Shibata:2009ad,Shibata:2010wz}), we here add a
  small perturbation to the initial data in order to trigger an instability more
  rapidly (if one exists). This perturbation introduces an additional constraint
  violation at a level well below that due to the discretization but we
  significantly mitigate this effect by localizing the perturbation well within
  the horizon.}; confining the fluctuation within the horizon (by choosing
$R_{0}\simeq 0$) will however produce only a weak gravitational wave
signal. Thus, we here choose to monitor quantities that describe the horizon
deformation, as explained in the next section. We find that our results,
described below, can also be used to understand the gravitational wave signal at
large distances.  Note also that for our choices of perturbation amplitudes $A$,
when looking at, for instance, the Hamiltonian constraint violation, we see no
noticeable differences when comparing with non-perturbed cases.

\section{Diagnostics}
\label{sec_diagnosis}

We analyze the result of our numerical investigations using
the following quantities. (i) The (coordinate invariant) horizon ``areal''  radius
\begin{equation}
\label{eq:rar}
\rar = \sqrt{r_H^2 + a^2} \,,
\end{equation}
where $r_H \equiv M + \sqrt{M^2 - a^2 - Q^2}$. (ii) The ratio between the polar and equatorial horizon circumferences
\begin{equation}
\label{eq:circ-ratio}
\mathcal{C}_{p} / \mathcal{C}_{e} = \frac{r_H^2}{\pi(r_H^2 + a^2)} 
    \int_{0}^{\pi} \sqrt{1 + \frac{a^2}{r_H^2} \cos^2 \theta  } \ d\theta \,,
\end{equation}
which, for known $M$ and $Q$, allows one to determine $a$.
Finally, we (iii) quantify the ``strength'' of the bar-mode perturbation in terms of the following distortion parameters~\cite{Saijo:2000qt,Franci:2013mma}
\begin{equation}
\label{eq:eta}
\eta_{+}  \equiv \frac{I^{xx} - I^{yy}}{I^{xx} + I^{yy}} \,, \qquad \eta_{\times} \equiv \frac{2I^{xy}}{I^{xx} + I^{yy}} \,,
\end{equation}
%
where
\begin{equation}
\label{eq:Iij}
I^{ij} = \int_{H} d^3x \sqrt{\gamma} x^i x^j
\end{equation}
is the quadrupole moment of the apparent horizon. We also compute the radiation from
the system by computing the Newman-Penrose scalar $\Psi_4$ at distances far from the BH.
We have found, however, that the behavior of $\{\eta_{+},\eta_{\times}\}$ is better suited to
analyze the response of the near BH region to the perturbations---which, as discussed
in the previous section, are initially concentrated in that region. This is a natural observation as the
BH potential barrier essentially traps the induced perturbations in the BH's vicinity.

To monitor the evolution, we compute the relative difference of both the areal radius $\rar$ of the apparent horizon and the measured BH spin to the known analytic value  
\begin{equation}
\label{eq:rel-err}
\delta[f(t)] \equiv \max_{t>5M} \frac{|f(t) - f^0|}{f^0} \,,
\end{equation}
where $f^0$ is the analytic value.
We choose to evaluate the maximum from $t\simeq 5M$ onward to remove possible large fluctuations due to our 
initial perturbation~\eqref{eq:pert}. Finally we monitor convergence of the solution via standard numerical
analysis and the behavior of the constraints to ensure truncation errors remain small throughout the simulation's time span.

\section{Numerical results}
\label{sec:numerical_results}

We numerically integrate the Einstein-Maxwell system using fourth-order spatial
discretization with the \textsc{Lean} code~\cite{Sperhake:2006cy}. This code is based on the
\textsc{Cactus} Computational toolkit~\cite{cactus}, the \textsc{Carpet} mesh
refinement package~\cite{Schnetter:2003rb,carpet} and uses
\textsc{AHFinderDirect} for tracking apparent
horizons~\cite{Thornburg:2003sf,Thornburg:1995cp}. \textsc{Lean} uses the BSSN
formulation of the Einstein equations~\cite{Shibata:1995we,Baumgarte:1998te}
with the moving puncture method~\cite{Campanelli:2005dd,Baker:2005vv}. We refer
the interested reader to Ref.~\cite{Sperhake:2006cy} for further details on the
numerical methods, and to~\cite{Zilhao:2013nda} for the tests performed with the Einstein-Maxwell implementation.

\begin{figure}[htbp]
\centering
\includegraphics[width=0.45\textwidth]{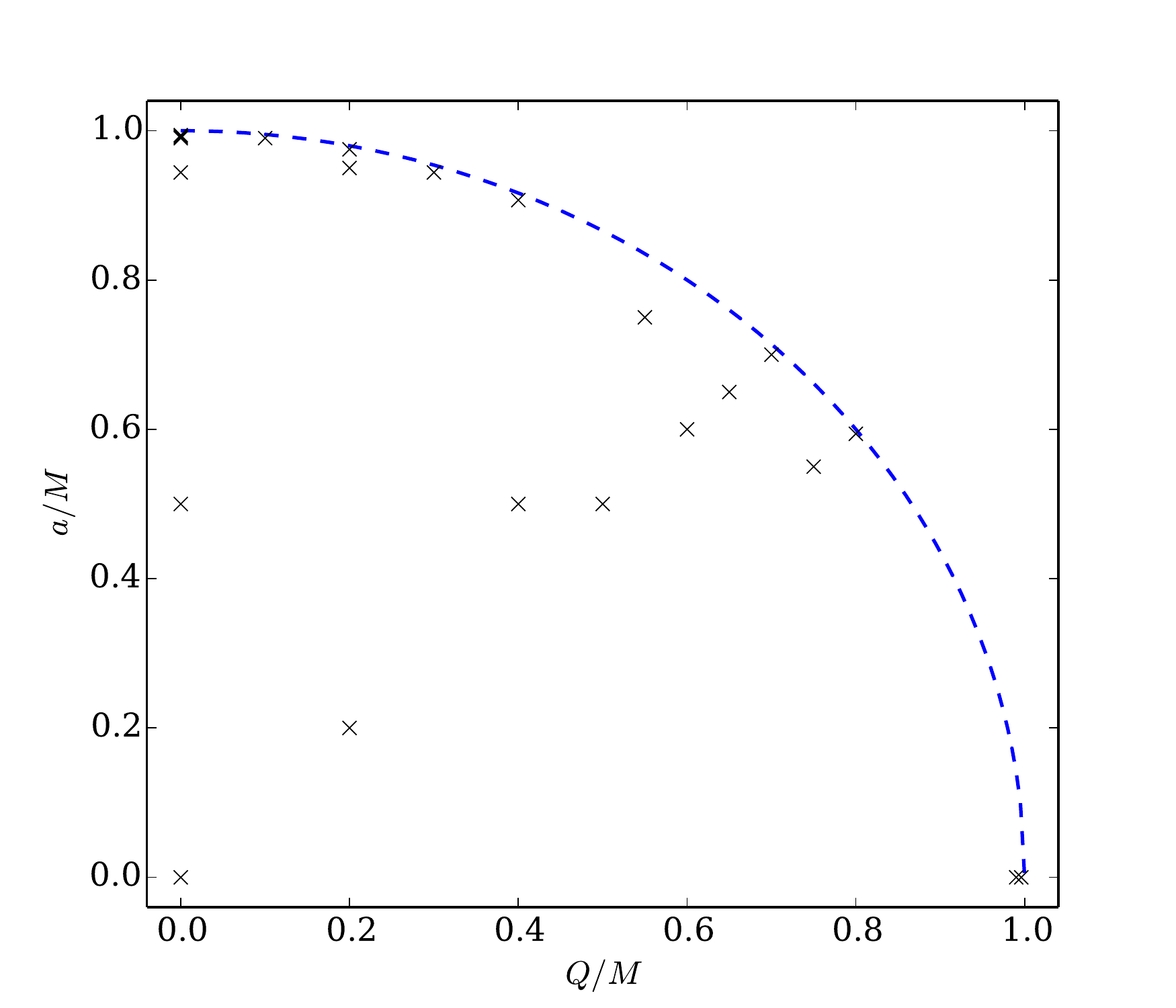}
\caption[]{The simulations performed in this work are displayed as crosses in the parameter space spanned by the rotation parameter $a$ and the charge $Q$.
The dashed blue line shows the extremal limit $a=a_{\rm max}$.
\label{fig:exclusion_plot} }
\end{figure}
We evolve the Einstein-Maxwell system of equations~(\ref{eq:EFE})
for several different charge and spin
values until $t\simeq 120M$ and  monitor
both the areal radius $\rar$ and the polar to equatorial horizon circumferences ratio.
As a practical measure, we consider a configuration to be \emph{stable} if: (i) during the course of
the numerical evolution, the BH areal radius and spin [the latter inferred
through Eq.~(\ref{eq:circ-ratio})] vary by less than a few
percent---consistent with the perturbation---with respect to the analytic value and (ii),
their time-dependence show an attenuating behavior. 
For visual guidance of the parameter space explored in this work, we display in
Fig.~\ref{fig:exclusion_plot} the extremality curve $a=a_{\rm max} \equiv
\sqrt{M^2-Q^2}$ together with the distribution of spin and charge values of the
simulations performed. The simulations performed include several configurations close
to extremality plus additional ones far from this regime for comparison purposes.

\begin{table*}[tbhp]
  \centering
  \caption{List of simulations performed with parameters used, where $a_{\rm max}\equiv \sqrt{M^2-Q^2}$. 
    The error reported was measured according to Eq.~(\ref{eq:rel-err}).
    For simulations with $a\geq0.99a_{\rm max}$, the numerical grid structure used (in the notation of Sec.~II~E
    of~\cite{Sperhake:2006cy}) was the following $\{(256,176,64,32,16,8,4,2,1,0.5,0.125), M/512\}$.
    \label{tab:runs}}
  
\begin{tabular*}{\textwidth}{@{\extracolsep{\fill}}lcccccc}
\hline 
\hline
Run  & 	  	  $a/M$ & 	  $Q/M$ &    $A$ &      $a/a_{\rm max}$  &    \% $\delta(\rar)$ &    \% $\delta (a)$ \\
\hline
\verb|a0.0_q0.0_A0.005|  & 	 0    & 	 0      &        0.005  & 	 0     & 	 0.00317 & 	 N.A. \\ 
\verb|a0.5_q0.0_A0.01|   &       0.5  & 	 0      & 	 0.01  & 	 0.5  & 	 0.056 & 	 1.74 \\ 
\verb|a0.944_q0.0_A0.0005|   & 	 0.944  & 	 0  & 	 0.0005  & 	 0.944  & 	 0.0527 & 	 0.0646 \\ 
\verb|a0.99_q0.0_A0.003|  &   	 0.99  & 	 0      & 	 0.003  & 	 0.99  & 	 0.759 & 	 0.327 \\ 
\verb|a0.992_q0.0_A0.0005|  &  	 0.992  & 	 0      & 	 0.0005  & 	 0.992  & 	 0.209 & 	 0.0678 \\ 
\verb|a0.994_q0.0_A0.0005|  & 	 0.994  & 	 0      & 	 0.0005  & 	 0.994  & 	 0.851 & 	 0.258 \\ 
\verb|a0.990_q0.1_A0.0005|  &  	 0.99  & 	 0.1  & 	 0.0005  & 	 0.995  & 	 2.92 & 	 1.05 \\ 
\verb|a0.2_q0.2_A0.02|      &     0.2  & 	 0.2  & 	 0.02  & 	 0.204  & 	 0.0404 & 	 1.36 \\ 
\verb|a0.95_q0.2_A0.005|    &  	 0.95  & 	 0.2  & 	 0.005  & 	 0.97  & 	 0.347 & 	 0.27 \\ 
\verb|a0.975_q0.2_A0.0005|  &  	 0.975  & 	 0.2  & 	 0.0005  & 	 0.995  & 	 2.7   & 	 0.938 \\ 
\verb|a0.944_q0.3_A0.0005|  &  	 0.944  & 	 0.3  & 	 0.0005  & 	 0.99  & 	 0.0403 & 	 0.0169 \\ 
\verb|a0.5_q0.4_A0.0|       &     0.5  & 	 0.4  & 	 0        & 	 0.546  & 	 0.0027 & 	 0.00652 \\ 
\verb|a0.907_q0.4_A0.0005|  &  	 0.907  & 	 0.4  & 	 0.0005  & 	 0.99  & 	 0.0596 & 	 0.026 \\ 
\verb|a0.5_q0.5_A0.02|      &    0.5  & 	 0.5  & 	 0.02  & 	 0.577  & 	 0.0292 & 	 0.154 \\ 
\verb|a0.75_q0.55_A0.005|   &  	 0.75  & 	 0.55  & 	 0.005  & 	 0.898  & 	 0.0157 & 	 0.067 \\ 
\verb|a0.6_q0.6_A0.005|     & 	 0.6  & 	 0.6  & 	 0.005  & 	 0.75  & 	 0.00356 & 	 0.0329 \\ 
\verb|a0.65_q0.65_A0.013|   &  	 0.65  & 	 0.65  & 	 0.013  & 	 0.855  & 	 0.00775 & 	 0.0363 \\ 
\verb|a0.7_q0.7_A0.005|     &  	 0.7  & 	 0.7  & 	 0.005  & 	 0.98  & 	 0.162 & 	 0.1 \\ 
\verb|a0.55_q0.75_A0.005|   &  	 0.55  & 	 0.75  & 	 0.005  & 	 0.832  & 	 0.00493 & 	 0.0195 \\ 
\verb|a0.594_q0.8_A0.0005|  & 	 0.594  & 	 0.8  & 	 0.0005  & 	 0.99  & 	 0.139 & 	 0.0601 \\ 
\verb|a0.0_q0.99_A0.0005|  &     0      & 	 0.99  & 	 0.0005  & 	 0     & 	 0.0119 & 	 N.A. \\ 
\verb|a0.0_q0.996_A0.0|     &  	 0      & 	 0.996  & 	 0       & 	 0     & 	 0.0329 & 	 N.A. \\ 
\hline
\hline
\end{tabular*}

\end{table*}
In Table~\ref{tab:runs} we list the simulations performed with the corresponding
physical parameters used.  Note that, except for two instances (runs \verb|a0.990_q0.1_A0.0005| and
\verb|a0.975_q0.2_A0.0005|, both of these being cases where $a=0.995a_{\rm
  max}$), the relative variation in $\rar$ is always smaller than $1\%$ (and for
most cases even smaller than $0.1\%$), which gives us confidence in the accuracy
of our numerical evolution since these are consistent with corresponding results for the
Schwarzschild case ($a=0=Q$).
The larger variation observed in the two mentioned
cases (and, to a lesser degree, also in the \verb|a0.992_q0.0_A0.0005| and
\verb|a0.994_q0.0_A0.0005| runs) is due to a small but steady growth in
$\rar$ observed from $t\sim 80M$ onward. We saw similar behaviour in other simulations accompanied by a steady increase of
the Hamiltonian constraint violations with time. In all such circumstances, this
behaviour was successfully cured with an increase in the numerical resolution
used. We believe that this is happening in all aforementioned cases: the growth
in the measured horizon area is merely telling us that more resolution is needed
should we want to accurately evolve such near-extremal configurations ($a
\gtrsim 0.994 a_{\rm max}$) for longer times. The already very high resolution used in
such cases effectively limits our ability to do so, however.

\subsection{Nonlinear stability of Kerr-Newman spacetimes}

\begin{figure}[htbp]
\centering
\includegraphics[width=0.45\textwidth]{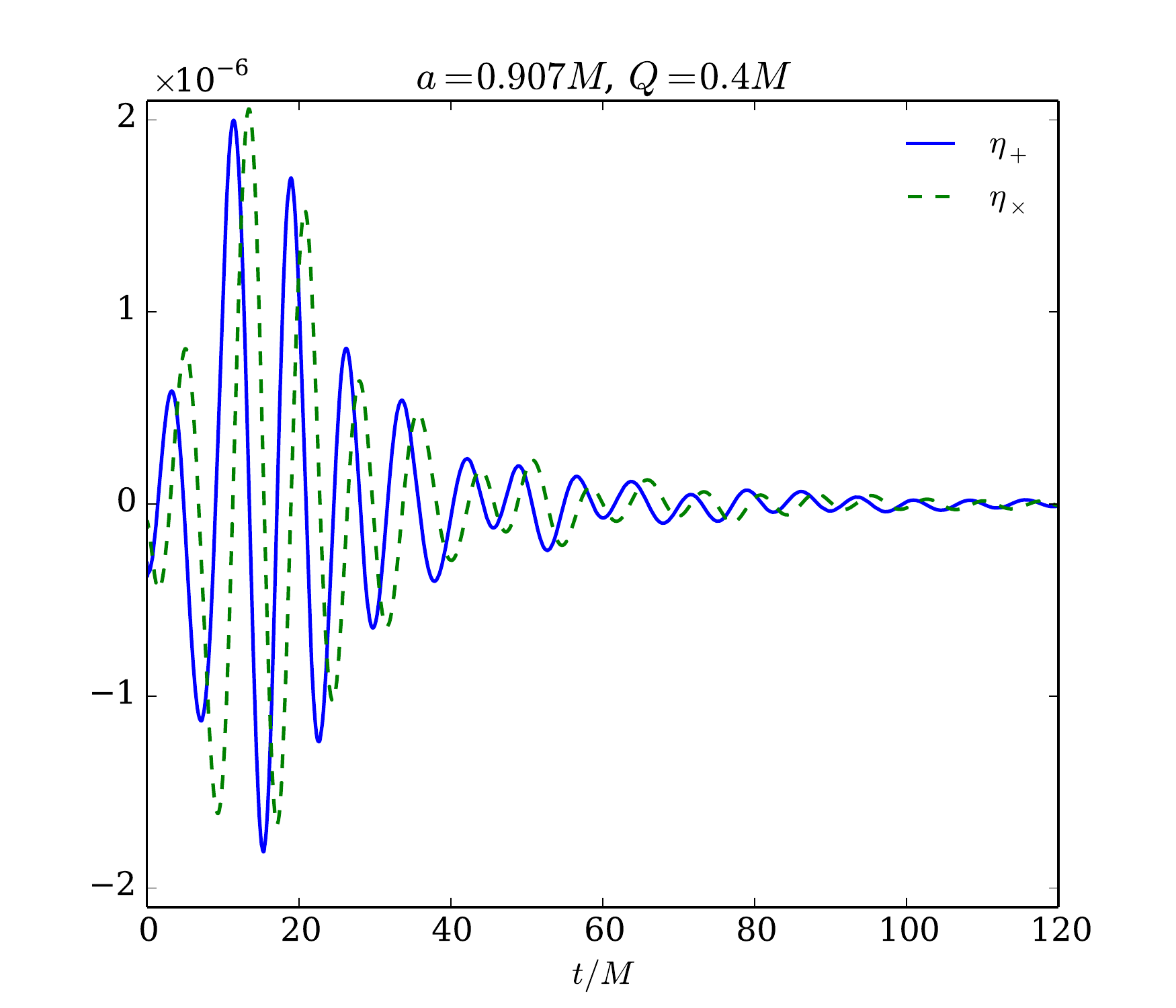}
\caption[]{Measured deformation parameters $\eta_{+,\times}$ as given from~(\ref{eq:eta}),
  as function of time for a simulation with $a=0.907M$, $Q=0.4M$,
  $A=5\times 10^{-4}$. \label{fig:eta} }
\end{figure}

\begin{figure}[htbp]
\centering
\includegraphics[width=0.45\textwidth]{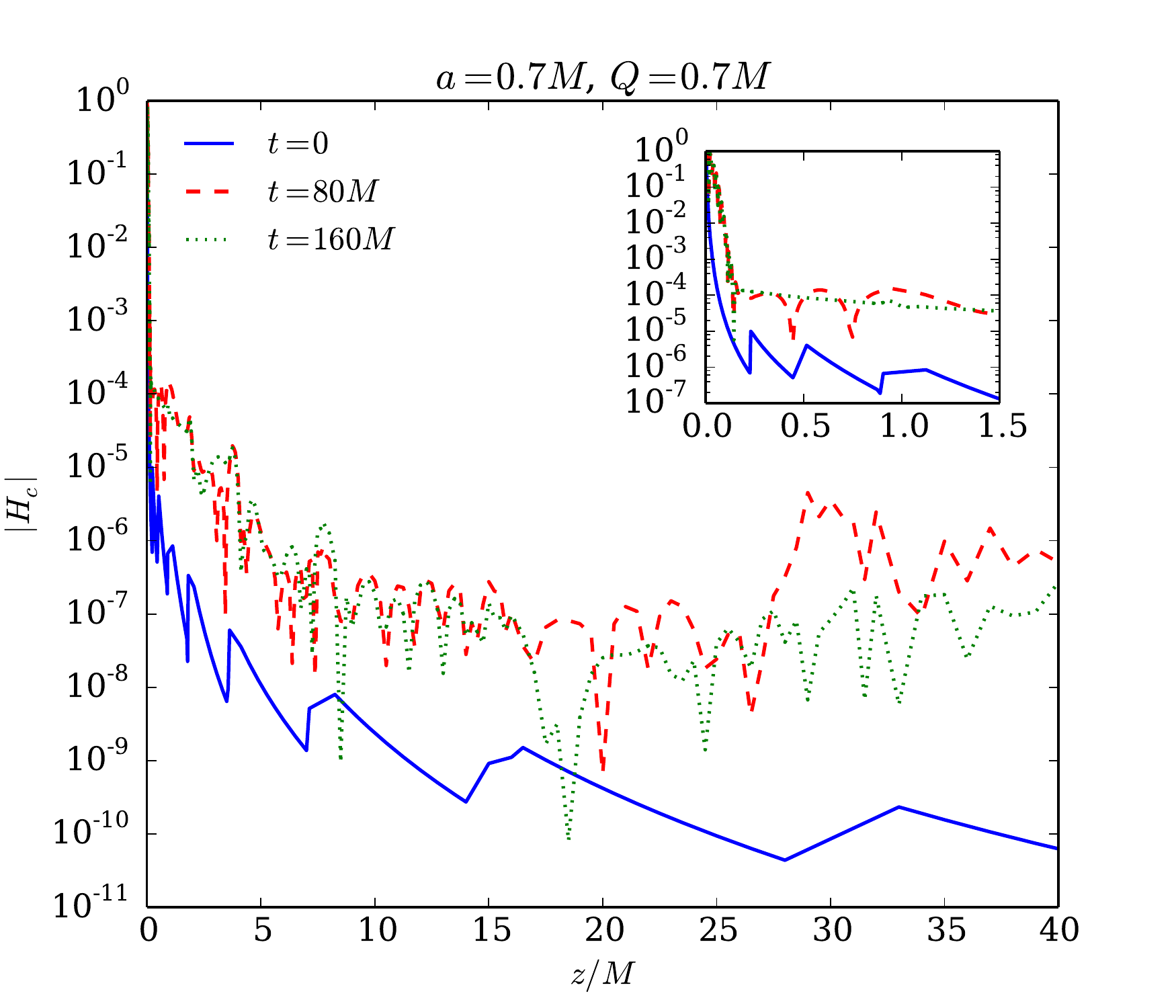}
\caption[]{Snapshots of the Hamiltonian constraint violation along the $z$-axis at taken at three different values of the evolution time for a simulation with $a=0.7M$, $Q=0.7M$, $A=0.005$. The inset shows the same data in a region close to the horizon [$R_H(t=0) \simeq 0.0707M$,
$R_H(t=160M) \simeq 0.16M$]. \label{fig:hc_a07_q07} }
\end{figure}

Figures~\ref{fig:eta}--\ref{fig:etaplus} summarize our results.
In Fig.~\ref{fig:eta} we plot the time evolution of the deformation parameters~\eqref{eq:eta} for a ``typical'' case corresponding to $(a/M,Q/M)=(0.907,0.4)$. 
The behavior of $\{\eta_{+},\eta_{\times}\}$ consists of a sum of damped sinusoids and decays away on timescales of order $100M$, consistent with linearized predictions for the ringdown timescale~\cite{Berti:2009kk}. For neutral or static BHs,
the ringing frequency and damping times of the fluctuations match well linearized calculations of quasinormal frequencies~\cite{Berti:2009kk}.
All our simulations display this same behavior: initial fluctuations are damped away. This is one of the main messages of our work:
for the parameters we studied, the KN geometry appears to be 
nonlinearly stable against such perturbations on the timescales examined herein---thus indicating any possible instability should have a secular growth associated to it. 

In Fig.~\ref{fig:hc_a07_q07}, we show the Hamiltonian constraint violations for a $(a/M,Q/M)=(0.7,0.7)$ example. Note that after a brief transient early on, the constraint
violation is not significantly growing in time, and that it bears the overall pattern observed for typical numerical BH evolutions.

\subsection{Universality of oscillation modes}
%
\begin{figure}[htbp]
\centering
\includegraphics[width=0.45\textwidth]{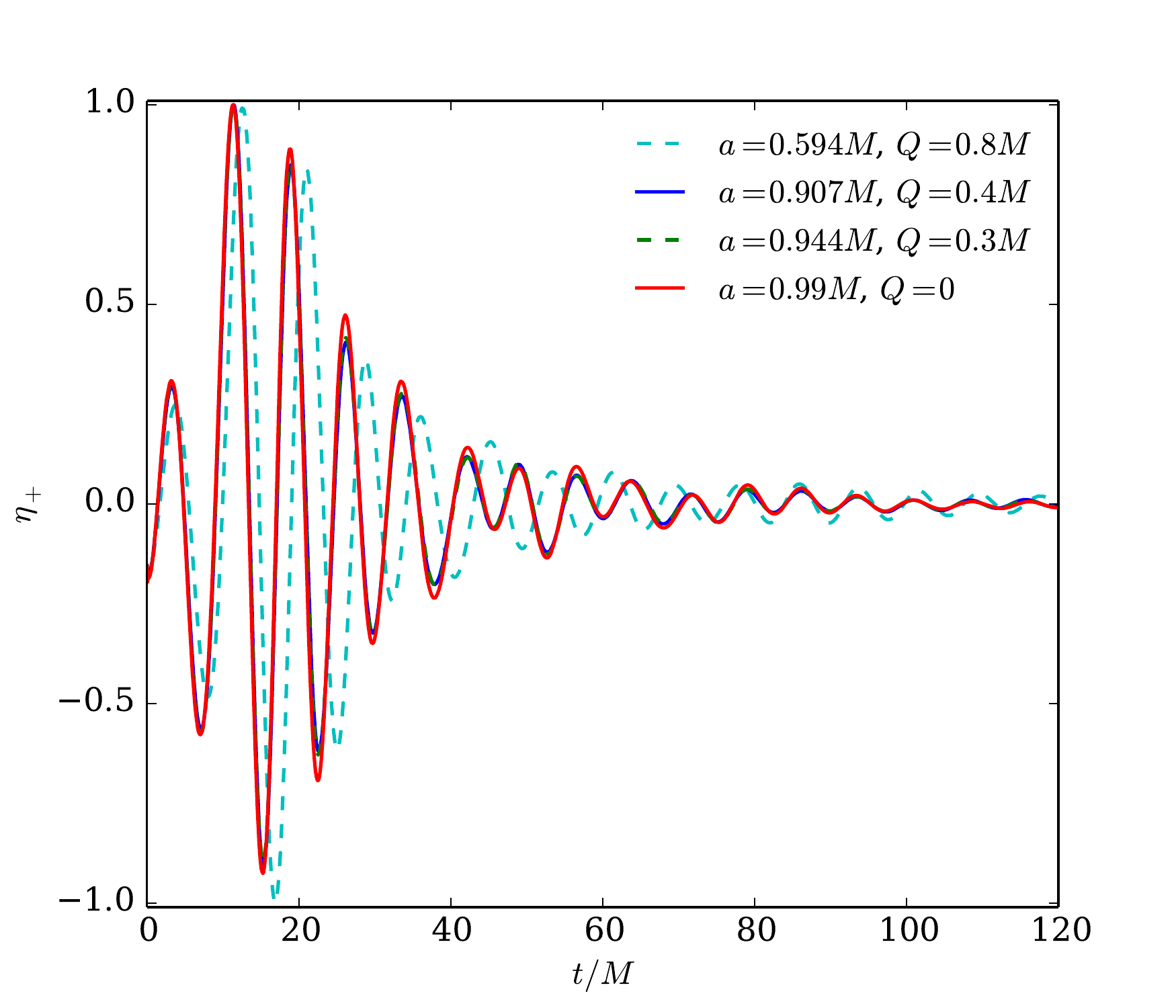}
\caption[]{Measured deformation parameter $\eta_{+}$ for several different
  simulations as function of time. All curves were normalized to their respective maximum
  amplitude. \label{fig:etaplus} }
\end{figure}
Our results indicate a surprising universal relation between the oscillation frequency and damping times of the fluctuations, 
namely that for large $a/Q$, spacetimes with the same $a/a_{\rm max}$ behave in a similar way.
This is summarized in Fig.~\ref{fig:etaplus} where we show the evolution of $\eta_+$ for three different values of $(a,Q)$
which share the same $a/a_{\rm max}=0.99$, and have $a/Q>1$. The lines corresponding to the different cases overlap almost perfectly. 
For comparison, another value of $(a,Q)$ with $a/a_{\rm max}=0.99$, but with $a/Q<1$ is exhibited, for which the curve is slightly displaced from the previous ones.  We note that if this agreement holds throughout the entire range
of charge and mass, this would imply that the characteristic or quasinormal frequencies of these BHs satisfy
\begin{equation}
\omega=\omega(a/\sqrt{M^2-Q^2})\,,\label{univ}
\end{equation}
which for small charge can also be written as $\omega=\omega(a/M+ay/M+...)$, where we defined $y \equiv 1-\sqrt{1-Q^2/M^2}$
following Refs.~\cite{Pani:2013ija,Pani:2013wsa}. This prediction was tested against linearized calculations in the slowly-rotating regime 
from~\cite{Pani:2013ija,Pani:2013wsa}, where frequencies are expressed as $M\omega=M\omega(a=0)+a\left(f_0+f_1y+...\right)$.
Translated into this notation, universality as described by Eq.~\eqref{univ} would imply that $f_0=f_1$ for both the real and imaginary components,
which is to very good precision the result presented in Table~I of~\cite{Pani:2013ija,Pani:2013wsa} for $l=2$ modes.

While such universality seems to hold only for quadrupolar modes (and again, the linearized calculations of Refs.~\cite{Pani:2013ija,Pani:2013wsa} are also consistent with universality for $l=2$ only), the mere existence of such property is intriguing and adds to the isospectrality found
in linearized studies~\cite{Pani:2013ija,Pani:2013wsa}.

Such universality is not an artifact of horizon-deformation measures. Our results indicate that the gravitational-wave signal at large distances (in particular the $l=m=2,4$ components of the scalar $\Psi_{4}$)  shares the same characteristics.

Recently, an analytical formalism to compute the quasinormal mode spectra of (weakly) charged black holes has been 
introduced~\cite{Mark:2014aja}.
With it, the extent of this seemingly universal behavior can be scrutinized. 
This has confirmed such behavior for large spin values (see also~\cite{Hod:2014uqa}), but it degrades 
considerably at low ones~\cite{markhuan}.

\section{Conclusions}
\label{sec:final}

In this paper we have used the techniques developed in~\cite{Zilhao:2012gp,Zilhao:2013nda} for performing BH evolutions in Einstein-Maxwell theory to study the nonlinear stability of the Kerr-Newman BH for a variety of parameters and, in particular, for rapidly spinning BHs. On the timescales explored here (a few hundred $M$), we have seen no evidence for instabilities in any of the simulations performed. We are able to measure the spin of the BH with high accuracy (see Table~\ref{tab:runs}), which varies only within the expected margin for numerical error, indicating that the solution is stable. 

In order to trigger potential instabilities, we have considered an initial perturbation of a particular type: a bar-mode perturbation in the metric coefficients.  We do not expect, however, that other types of qualitatively different initial perturbations---like Brill or Teukolsky waves (see e.g.~\cite{Hilditch:2013cba}, for a recent study using this type of initial data in moving puncture gauge) will give different results; otherwise, an instability would appear to require very specific perturbations neither contained in our bar mode nor in the numerical noise of the initial data.
As such, our nonlinear analysis reinforces previous linear results~\cite{Pani:2013ija,Pani:2013wsa,Civin:2014bha} on the stability of the nonextremal KNBH. This contrasts with the instability found for extremal KNBHs~\cite{Aretakis:2012ei,Reiris:2013efa}. Thus, the latter, albeit continuously connected to nonextremal KNBHs in parameter space, seem qualitatively disconnected in terms of physical properties.

Our results have also uncovered, in the large rotation regime, a new class of universality for the quadrupolar quasinormal modes of these BHs:
they depend solely on the combination $a/a_{\rm max}$, a feature which had been observed previously in the perturbative regime
of slow-rotation. The significance of such results is unclear, but together with the isospectrality---observed also in the slow-rotation regime---hints at deeper relations at work also in rotating and charged geometries.

\begin{acknowledgments}

We would like to thank Paolo~Pani, Zachary~Mark and Huan~Yang for useful discussions.
M.Z.\ is supported by NSF grants OCI-0832606, PHY-0969855, AST-1028087, and PHY-1229173.
V.C.\ acknowledges financial support provided under the European Union's FP7 ERC Starting Grant ``The dynamics of black holes:
testing the limits of Einstein's theory'' grant agreement no.\ DyBHo--256667. 
L.L.\ acknowledges support by NSERC through a Discovery Grant and CIFAR.
U.S.\ acknowledges support by
the FP7-PEOPLE-2011-CIG CBHEO Grant No. 293412,
the STFC Grant No. ST/I002006/1,
the XSEDE Grant No. PHY-090003 by the National Science Foundation,
the COSMOS Shared Memory system at DAMTP, University of Cambridge, operated on
behalf of the DiRAC HPC Facility and funded by BIS National E-infrastructure
capital Grant Nos.~ST/J005673/1 and ST/J001341/1
and STFC Grant Nos.~ST/H008586/1
and ST/K00333X/1, and
the Centro de Supercomputacion de Galicia (CESGA) under Grant No. ICTS-2013-249.
This research was supported in part by Perimeter Institute for Theoretical Physics. 
Research at Perimeter Institute is supported by the Government of Canada through 
Industry Canada and by the Province of Ontario through the Ministry of Economic Development 
$\&$ Innovation.
This work was supported by the NRHEP 295189 FP7-PEOPLE-2011-IRSES Grant, and by FCT-Portugal through projects
PTDC/FIS/116625/2010, CERN/FP/123593/2011 and the IF program.
Computations were performed on the ``Baltasar Sete-Sois'' cluster at IST, the ``Blafis'' cluster at Universidade de Aveiro, the NICS Kraken Cluster,
the SDSC Trestles Cluster, Cambridge's COSMOS,
on the ``venus'' cluster at YITP, and CESGA's Finis Terrae.

\end{acknowledgments}

\bibliographystyle{myutphys}
\bibliography{num-rel}

\end{document}